\documentclass[11pt]{article}  
\usepackage[centertags]{amsmath}
\usepackage{amsfonts} 
\usepackage{amssymb} 
\usepackage{amsthm}
\makeatletter \def\section{\@startsection {section}{1}{\z@}{-3.5ex
plus -1ex minus -.2ex}{2.3ex plus .2ex}{\large\bf}}
\def\subsection{\@startsection{subsection}{2}{\z@}{-3.25ex plus -1ex
minus -.2ex}{1.5ex plus .2ex}{\normalsize\bf}}
\newcommand{\braket}[2]{\langle{#1}|{#2}\rangle}
\newcommand{\be}{\begin{equation}} \newcommand{\ee}{\end{equation}}
\newcommand{\bea}{\begin{eqnarray}} \newcommand{\eea}{\end{eqnarray}}
\makeatletter \@addtoreset{equation}{section} \makeatother

\def\ZZ{\relax\ifmmode\mathchoice {\hbox{\cmss Z\kern-.4em
Z}}{\hbox{\cmss Z\kern-.4em Z}} {\lower.9pt\hbox{\cmsss Z\kern-.4em
Z}} {\lower1.2pt\hbox{\cmsss Z\kern-.4em Z}}\else{\cmss Z\kern-.4em
Z}\fi} \def\IR{\relax{\rm I\kern-.18em R}}

\begin{document}
\begin{titlepage}
\rightline{NORDITA-2001/92 HE}  \rightline{DSF-43/2001}
\vskip 3.0cm
%\vskip 0.8cm 
\centerline{\LARGE \bf Fractional Branes and $\mathbf{\mathcal{N}=1}$
Gauge Theories}
\vskip 1.4cm  \centerline{\bf M. Bertolini $^a$, P. Di Vecchia $^a$,
G. Ferretti $^b$ and R. Marotta $^c$}
\vskip .8cm \centerline{\sl $^a$ NORDITA, Blegdamsvej 17, DK-2100
Copenhagen \O, Denmark}
\vskip .4cm  \centerline{\sl $^b$ Institute for Theoretical Physics -
G\"oteborg University and} \centerline{\sl Chalmers University of
Technology, 412 96 G\"oteborg, Sweden}
\vskip .4cm  \centerline{\sl $^c$ Dipartimento di Scienze Fisiche,
Universit\`a di Napoli and INFN, Sezione di Napoli}  \centerline{\sl
Via Cintia - Complesso Universitario M. Sant' Angelo I-80126 Napoli,
Italy}
\vskip 2cm
\begin{abstract}
We discuss fractional D3-branes on the orbifold $\mathbb{C}^3 /
\mathbb{Z}_2\times \mathbb{Z}_2$. We study the open and the closed
string spectrum on this orbifold. The corresponding ${\cal N}=1$
theory on the  brane has, generically, a $U(N_1) \times U(N_2) \times
U(N_3) \times U(N_4)$ gauge group with matter in the bifundamental. In
particular,  when only one type of brane is present, one obtains pure
${\cal N}=1$  Yang-Mills. We study the coupling of the branes to the
bulk fields and present the corresponding supergravity solution, valid
at large distances. By using a probe analysis, we are able to obtain
the Wilsonian  $\beta$-function for those gauge theories that possess
some chiral  multiplet.  Although, due to the lack of moduli,  the
probe technique is not directly applicable to the case of pure ${\cal
N}=1$ Yang-Mills, we point out that the same formula gives the
correct result also for this case.
\end{abstract}

\end{titlepage}

\renewcommand{\thefootnote}{\arabic{footnote}}
\setcounter{footnote}{0} \setcounter{page}{1}

%%%%%%%%%%%%%%%%%%%%%%%%%%%%%%%%%%%%%%%%%%%%%%%%%%%%%%%%%%%%%%%%%%%%%%%%%%%
\tableofcontents
\vskip 1cm

%%%%%%%%%%%%%%%%%%%%%%%%%%%%%%%%%%%%%%%%%%%%%%%%%%%%%%%%%%%%%%%%%%%%%%%%%%
%%%%%%%%%%%%%%%%%%%%%%%%%%%%%%%%%%%%%%%%%%%%%%%%%%%%%%%%%%%%%%%%%%%%%%%%%%
\section{Introduction}
Fractional branes
\cite{Douglas:1997xg,Douglas:1997de,Diaconescu:1998br,marco1} provide   
a useful way to construct gauge theories with reduced supersymmetry
in string theory. In particular, for theories on orbifold, from the
asymptotic behaviour of the fields belonging to the twisted sector,
one can read off various quantities of the guage theory. For instance,
one was able to obtain the perturbative $\beta$-function of pure
${\cal N}=2$ Yang--Mills theory from this perspective
\cite{Bertolini:2000dk}.

The basic idea involved in this type of computations is that of
probing the gauge theory living on a stack of branes by putting a
single brane of similar kind next to them. The picture emerging from
the gauge theory is that one has higgsed the original theory by giving
a v.e.v. to the  matter multiplet describing the position of the probe
brane. The gauge theory on the brane probe becomes free and its
coupling constant stops running.  By reading off such frozen coupling
one obtains an algebraic relation for the running couplings at the
moment of the breaking and can thus reconstruct the
$\beta$-functions. The determination of the coupling on the probe is
obtained by looking at the Born--Infeld action and it is essentially
given by the (pull-back of) the twisted fields obtained from
supergravity \cite{Klebanov:2000rd}.

The above technique is related to the so called gauge/gravity
correspondence although, when discussing non conformal theories, one
is generically unable (with noticeable exceptions~\cite{KLEBA3}) to
obtain a singularity free gravity dual. In particular, fractional
branes will always give rise to singular gravity backgrounds but one
has nevertheless been able to obtain perturbative ${\cal N}=2$
$\beta$-functions in this way
\cite{Bertolini:2000dk,Polchinski:2000mx,Billo:2001vg,Bertolini:2001qa}
and it is not inconceivable that instanton corrections may also be
computed.% (see \cite{fu} for some discussion on this point).

In this paper we generalize these techniques to ${\cal N}=1$ gauge
theories.  In order to use the probe analysis, we need a theory with
some chiral  multiplets, and the orbifold $\mathbb{C}^3/(\mathbb{Z}_2
\times \mathbb{Z}_2)$ provides one of the simplest examples. There are
four different types of fractional D3-branes in such orbifold
\cite{diacgom},  none of which is free to  move separately. However,
we will see that they can move in pairs along certain orbifold
directions due to the presence of chiral multiplets and this allows us
to  construct a bound state probe and to derive a certain linear
combination of $\beta$-functions.  The bound state probe moving in one
of the three complex directions is  nothing but a fractional brane of
the $\mathbb{C}^2/\mathbb{Z}_2$ theory  of~\cite{Bertolini:2000dk}
with the scalars describing its position in relation to the orbifold
point.  Thus, when we are probing ${\cal N}=1$ theories, the low
energy theory on the probe is a (free) $U(1)$ ${\cal N}=2$ theory.

The computation of the $\beta$-function proceeds in a way that is very
similar to that of~\cite{Bertolini:2000dk}. The boundary state
technique allows one to obtain the (properly normalized) coupling of
the  brane to the twisted fields. Thus, the brane acts as a source for
such fields and their logarithmic behaviour is re-interpreted, via the
Born--Infeld action, as the running coupling.  Because of the need for
chiral multiplets in the bifundamental, the  gauge group under study
contains at least two simple factors, each one coming with its own
coupling constant.  Since the gauge symmetry breaking in the ${\cal
N}=1$ case is slightly  more involved than that for ${\cal N}=2$, the
probe analysis allows one to  obtain only a linear combination of the
$\beta$-functions for these couplings.  However, we show that the
formul{\ae} obtained give the correct answer even in the case of pure
${\cal N}=1$ Yang--Mills with only one $U(N)$ gauge group although,
strictly speaking, the probe analysis is not  applicable there. We
believe that it should be possible to justify this  result with a
refinement of the probe analysis.

It is perhaps worth emphasizing that the results for the
$\beta$-functions presented here are valid near the gaussian UV fixed
point, where the anomalous dimensions vanish. This is justified here
because we start with a renormalizable theory which admits a continuum
limit.  This should not be confused with the case discussed in
\cite{Klebanov:1998hh,KLEBA3}, where the results for the
$\beta$-functions concern the theory at a  non-gaussian IR fixed
point, and contain some non perturbative information encoded in the
anomalous dimension of the fields via the exact relation of
\cite{Shifman:1986zi}.

The paper is organized as follows. In section 2, we discuss the open
and closed string spectrum for fractional branes at the orbifold
$\mathbb{C}^3/(\mathbb{Z}_2 \times \mathbb{Z}_2)$ and, using the
technique of boundary state, we compute the coupling of the various
bulk fields to the brane and construct the boundary action.  In
section 3, we obtain the gravity solution corresponding to the most
generic combination of fractional branes. The part of the solution
that is relevant to the computation of the $\beta$-functions is the
behaviour of the  twisted fields. Although the solution for the metric
is singular (as it was in the ${\cal N}=2$ case), the probe analysis
is justified at large distances. In section 4, we will use these
results to obtain the $\beta$-functions of the gauge theory. We will
also justify the probe analysis from the gauge theory point of view by
studying the higgsing of the theory. Moreover, noticing that brane
probes become tensionless before reaching the singularity, we also
conclude that the short-distance region of spacetime is out of reach
of the supergravity solution while the singularity is excised. This
implies that we cannot probe the IR of the gauge theory, but this is
not relevant in computing the perturbative $\beta$-function. Finally,
we will show that our formul{\ae} also correctly reproduce the
$\beta$-function for the pure ${\cal N}=1$ theory when extrapolated
outside of the regime  of validity of the probe analysis.

%%%%%%%%%%%%%%%%%%%%%%%%%%%%%%%%%%%%%%%%%%%%%%%%%%%%%%%%%%%%%%%%%%%%%%%%%%
%%%%%%%%%%%%%%%%%%%%%%%%%%%%%%%%%%%%%%%%%%%%%%%%%%%%%%%%%%%%%%%%%%%%%%%%%%
\section{Regular and fractional branes on  
$\mathbb{C}^3/(\mathbb{Z}_2 \times \mathbb{Z}_2)$}
\label{gspec}

In this section we consider both regular and fractional D3-branes of
type IIB string theory on the orbifold $\mathbb{R}^{1,3} \times
\mathbb{C}^3/(\mathbb{Z}_2 \times \mathbb{Z}_2)$,  we study the
spectrum of the massless open string states having their end-points
attached to them and, after constructing the boundary state encoding
their properties, we determine their boundary action and  the large
distance behaviour of the classical solution corresponding to them.

The group $\mathbb{Z}_2 \times \mathbb{Z}_2$ has four elements: the
identity $e$, the generators of the two $\mathbb{Z}_2$ that we denote
with $h_1$ and $h_2$ and their product, denoted by $h_3 =h_1 h_2$. By
taking the orbifold directions to be along $x^4,\dots,x^9$  and
introducing complex coordinates $(z_1, z_2,z_3)\in \mathbb{C}^3$
defined by:
\begin{equation}
z_1 = x^4 + i x^5~~,~~z_2 = x^6 +i x^7~~,~~ z_3 = x^8 + i x^9
\label{zzz67}
\end{equation}
the action of the orbifold group on the complex coordinates  $z_i$
($i=1,2,3$)  can be defined as:
\begin{table} [ht]
\label{orbac}
\begin{center}
\begin{tabular}{c||c|c|c|}
    & $z_1$ & $z_2$ & $z_3$ \\ \hline  \hline $e$    &$z_1$ & $z_2$ &
$z_3$ \\  \hline $h_1$  &$z_1$ & $-z_2$ & $-z_3$ \\  \hline $h_2$
&$-z_1$ & $z_2$ & $-z_3$ \\  \hline $h_3$  &$-z_1$ & $-z_2$ & $z_3$ \\
\hline \end{tabular}
\caption{\small The action of the orbifold generators on
$\mathbb{C}^3$.}
\end{center}
\end{table}    

\noindent
We are interested in studying D3-branes which are transverse to the
orbifold, namely with world-volume directions $x^{\alpha}$, with
$\alpha =0,1,2,3$.

The low energy closed string sector of the orbifold $\mathbb{C}^3/
(\mathbb{Z}_2 \times \mathbb{Z}_2 )$ consists of an untwisted sector
and three twisted sectors, corresponding to zero modes of supergravity
fields dimensionally reduced on the three exceptional vanishing
two-cycles ${\cal C}_i$ ($i=1,2,3$) characterizing the orbifold, each
of them embedded in one of the three four-dimensional subspaces of
$\mathbb{C}^3$ \cite{Douglas:1999hq}. The three anti self-dual 2-form
$\omega_{2}^{i}$, dual to the cycles ${\cal C}_i$, are then completely
independent and normalized  as:
\begin{equation}
\int_{{\cal C}_i} \omega_2^j = \delta_i^j \quad , \quad  \int
\omega_2^i \wedge \omega_2^i = -\frac{1}{4}   \quad , \quad *_4 \,
\omega_{2}^i = -\omega_{2}^{i}
\label{ome56}
\end{equation}
where the index $*_4$ indicates the dual in the four-dimensional space
in which the two-cycle is embedded.

Let us start by analyzing regular branes, namely those D-branes which
are free to  move in the full transverse space. In order to study the
open string spectrum  living on a regular D3-brane, it is convenient
to consider the covering space where  together with the original brane
there are also its three images. If the D3-brane is  located at an
arbitrary point of the transverse six-dimensional space, the open
strings stretched between the brane and its images or between two of
its images correspond in general to massive states. But massless
states appear when we put the D3-brane at the orbifold fixed point
$z_1 = z_2 = z_3 =0$.  The generic  open string state is the product
of a Chan-Paton factor consisting of a $4 \times 4$ matrix describing
the open strings attached to the D3-brane and its images and of an
oscillator part. In particular a massless state of the NS sector has
the following form:
\begin{equation}
\label{os1}
\lambda \,\otimes \, \psi^M_{-1/2}|0, k >
\end{equation}
where $\lambda$ denotes the Chan-Paton factor and  $M=0,1,...,9$.  The
action of the generators of $\mathbb{Z}_2 \times \mathbb{Z}_2$ on the
Chan-Paton factors is defined as
\begin{equation}
\label{cp1}
\gamma(h)\,\lambda\,\gamma^{-1}(h) = \lambda' \quad \mbox{for}  \quad
h = e,\; h_1,\;h_2,\;h_3
\end{equation}
It is convenient to choose the matrices $\gamma(h)$ to be:
\begin{equation}
\gamma(e) = \mathbf{1} \otimes \mathbf{1} \;\; ,\;\; \gamma(h_1) =
\sigma_3 \otimes \mathbf{1} \;\;,\;\;  \gamma(h_2) = \mathbf{1}
\otimes \sigma_3 \;\; ,\;\; \gamma(h_3) = \sigma_3 \otimes \sigma_3
\label{regre8}
\end{equation}
where with $\otimes$ we denote the usual tensorial product.

Not all states in eq.(\ref{os1}) are allowed in an orbifold
theory. The only allowed states are those that are left invariant
under the combined action of the orbifold group on the oscillators and
the Chan-Paton factors.  In particular, if $M= \alpha$, the oscillator
part is left invariant by the action of the orbifold group.
Therefore, we must require that also the Chan-Paton part be left
invariant by the action of  $\gamma(h)$, namely $\lambda'=\lambda$  in
eq.(\ref{cp1}). On the other hand, since $\psi^{4,...,9}_{-1/2}|0, k
>$ can  be either even or odd under the action of the orbifold group,
the states surviving the orbifold projection are those with Chan-Paton
factors that are respectively even or odd under the action of the
orbifold group\footnote{Remember that  world-sheet  supersymmetry
requires that the orbifold group acts in the same way on the bosonic
and fermionic coordinates.}. By a careful analysis one finds the
following bosonic spectrum of massless states:
\begin{eqnarray}
%\begin{tabular}{c|c}
\underline{\mbox{Vectors:}}  && \lambda \times \psi^\alpha_{-1/2} | k>
\quad  \lambda = \{ \mathbf{1} \otimes \mathbf{1} \;,\; \sigma_3
\otimes \mathbf{1} \;,\;  \mathbf{1} \otimes \sigma_3 \;,\; \sigma_3
\otimes \sigma_3 \} \nonumber \\   \underline{\mbox{Scalars:}}  &&
\lambda \times \psi^{4,5}_{-1/2} | k> \quad \lambda = \{
\mathbf{1}\otimes i \sigma_2 \;,\; \mathbf{1}\otimes \sigma_1 \;,\;
\sigma_3 \otimes i \sigma_2 \;,\; \sigma_3 \otimes \sigma_1\}
\nonumber \\  && \lambda \times \psi^{6,7}_{-1/2} | k> \quad  \lambda
= \{ i \sigma_2 \otimes \mathbf{1}\;,\; \sigma_1 \otimes
\mathbf{1}\;,\; i \sigma_2 \otimes \sigma_3 \;,\; \sigma_1 \otimes
\sigma_3\}\nonumber \\  && \lambda \times \psi^{8,9}_{-1/2} | k> \quad
\lambda = \{ i \sigma_2 \otimes \sigma_1\;,\; i \sigma_2 \otimes i
\sigma_2\;,\; \sigma_1 \otimes i \sigma_2\;,\; \sigma_1 \otimes
\sigma_1 \} \nonumber
%\end{tabular}
\end{eqnarray}
Including also the fermionic spectrum obtained from the Ramond sector
one obtains 4 ${\cal{N}}=1$ gauge  and 12 chiral multiplets. The gauge
theory living on a regular D3-brane that, as we have seen, is
described by $4 \times 4$ Chan-Paton factors, has thus gauge group
$U(1)_1 \times U(1)_2 \times U(1)_3 \times U(1)_4$ and $12$ chiral
multiplets. It is convenient to use a basis in the space of the $4
\times 4$ Chan-Paton factors where each diagonal entry corresponds to
one of the $U(1)$ factors:
\begin{equation} 
A_\alpha =
\begin{pmatrix}
A_\alpha^1 & 0        & 0       & 0        \cr  0        & A_\alpha^2
& 0       & 0        \cr  0        & 0        &  A_\mu^3 & 0 \cr  0 &
0        &  0      & A_\alpha^4  \cr
\end{pmatrix}
\end{equation} 
A similar structure holds for the 12 chiral  multiplets, which, in the
above gauge field basis,  can be organized in three $4\times 4$
matrices given by
\begin{equation}
\label{sc1}
\Phi_1 = \begin{pmatrix}0 & a_1 & 0   & 0   \cr  b_1 & 0 & 0   & 0 \cr
                    0 & 0   & 0   & c_1 \cr  0   & 0   & d_1 & 0
                    \cr\end{pmatrix}\;,\; \Phi_2 = \begin{pmatrix}0 &
                    0   & a_2 & 0  \cr     0   & 0   & 0   & b_2\cr
                    c_2 & 0   & 0   & 0  \cr  0   & d_2 & 0 & 0
                    \cr\end{pmatrix}\;,\;  \Phi_3 = \begin{pmatrix}0 &
                    0   & 0   & a_3 \cr  0   & 0 & b_3 & 0   \cr  0 &
                    c_3 & 0   & 0   \cr  d_3 & 0   & 0   & 0
                    \cr\end{pmatrix}
\end{equation}
where $a_i,\cdots,d_i$ are each a chiral multiplet and we picked the
same complex structure as  in eq.(\ref{zzz67}).

The superpotential can be written as $W=\mathrm{tr}(\Phi_1[\Phi_2,
\Phi_3])$ and can be easily generalized to the non abelian case. To
avoid confusion, it is worth  noticing that, contrary to the quartic
superpotential  of~\cite{Klebanov:1998hh}, this superpotential is
renormalizable in the  UV. As remarked in the introduction, our
results are valid in that region,  where we can use the perturbative
expression for the $\beta$-functions  with all the anomalous
dimensions $\gamma\approx 0$.

Looking at the $U(1)$ charges of these multiplets we see that each of
them is charged with respect to \emph{two} gauge fields. This means
that the chiral multiplets transform in the bifundamental of any given
couple $I,J$ ($I,J=1,2,3,4$) of gauge groups and that each chiral
multiplet contributes to two different  $U(1)_I$. As a consequence
there are 6 chiral multiplets charged under a given gauge  group. If,
instead of only one, we have $N$ regular D3-branes, then the gauge
theory living on them is a supersymmetric ${\cal{N}}=1$ gauge theory
with gauge group $U_1(N)\times U_2(N)\times U_3(N)\times U_4(N)$ and
12 chiral multiplets, transforming  according to the fundamental
(anti-fundamental) representation of a given gauge group and  carrying
a flavor index in the fundamental (anti-fundamental) of one of  the
other 3 gauge groups. This theory,  as expected, is conformal. Indeed
for any of the four gauge groups the (Wilsonian) $\beta$-function
reads:
\begin{equation}
\beta_I = \frac{g_I^3}{(4\pi)^2}  [\underbrace{-\frac{11}{3} N +
\frac{2}{3} N}_{{Gauge\;multiplet}} + \underbrace{\frac{1}{6} 6N +
\frac{1}{3} 6N}_{{Chiral\;multiplets}}] = 0
\label{beta00}
\end{equation}
In conclusion we have seen that a regular brane is described by
Chan-Paton factors transforming according to a $4 \times 4$
representation of the discrete orbifold group $\mathbb{Z}_2 \times
\mathbb{Z}_2 $ and that the gauge theory living on it is a conformal
invariant theory.

It is known that a discrete abelian group, as the orbifold group
$\mathbb{Z}_2 \times \mathbb{Z}_2$, has only one-dimensional
irreducible representations. This means that a regular D3-brane must
be decomposable into more elementary objects, the fractional
D3-branes, whose Chan-Paton factors are indeed just numbers and
transform irreducibly under the orbifold group. The fact that the
representation of the orbifold group is reducible can be directly seen
from the explicit expression for the regular representation given in
eq.(\ref{regre8}) where all the $4 \times 4$ matrices are
diagonal. From them one can extract the four one-dimensional
irreducible representations of  $\mathbb{Z}_2 \times \mathbb{Z}_2$
corresponding to the four types of fractional branes. They read as
follows:
\begin{align}
&\gamma_1 (e) = +1& &\gamma_1 (h_1) = +1&  &\gamma_1 (h_2) = +1&
&\gamma_1 (h_3)  = +1&
\label{ga1} \\
&\gamma_2 (e) = +1& &\gamma_2 (h_1) = +1& &\gamma_2 (h_2) = -1&
&\gamma_2 (h_3)  = -1&
\label{ga2} \\
&\gamma_3 (e) = +1& &\gamma_3 (h_1) = -1& &\gamma_3 (h_2) = +1&
&\gamma_3 (h_3)  = -1&
\label{ga3} \\
&\gamma_4 (e) = +1& &\gamma_4 (h_1) = -1& &\gamma_4 (h_2) = -1&
&\gamma_4 (h_3)  = +1&
\label{ga4}
\end{align}
The first column is related to the coupling of the fractional branes
to the untwisted sector, while the other three columnss correspond to
the coupling of the fractional branes to the three twisted sectors
(with charge according to the corresponding $\pm$ sign). As  already
noticed, these three twisted sectors are related to the three
shrinking cycles ${\cal C}_i$ which are located at the orbifold fixed
point.  Since, according to the above table, fractional branes couple
to all twisted sectors, a given fractional D3-brane, whose transverse
space coincides with the orbifold directions, is stuck at the orbifold
fixed point and there is no way of moving it. The same is true if we
consider a bound state of an odd number of fractional D3-brane
types. On the other hand, if we consider a bound state of an even
number of fractional D3-brane types, it is possible to move it along
some directions of the six-dimensional  transverse space. Indeed, from
the above table, one can easily see that a bound state of four
fractional D3-branes, corresponding  to a regular D3-brane, can be
moved over the entire six-dimensional transverse space because the
twisted charges cancel and there is no coupling with any twisted
sector. For a bound state of  two fractional D3-brane types, the
charge under two twisted sectors cancels  and the bound state is
charged only under one twisted sector. Hence, the  bound state is free
to move along the two-dimensional space orthogonal to  the four
directions on which the one twisted sector is stuck. This observation
will be relevant in section 4, when we will discuss D-brane probes.

In terms of constituent fractional branes it is now easy to understand
the structure of the gauge group of $N$ regular D-branes as a product
of the  four $U(N)$ gauge groups we have found before. Each of the
$U(N)$ gauge fields corresponds to the massless open strings having
their end-points on one of the four fractional D3-branes constituting
the regular D3-brane. In particular, one could have a more general
gauge group $U(N_1 ) \times U(N_2 ) \times U(N_3 ) \times U(N_4 )$ if
we had considered a more general bound state of fractional D3-branes
consisting of $N_I$ fractional branes of each kind.

From the previous picture it follows that the gauge theory living on a
bound state made of two fractional D3-branes of different kind has
gauge group $U(1) \times U(1)$. As already noticed, such a D3-brane
can now move in the two-dimensional subspace of the entire
six-dimensional transverse space orthogonal to the four-dimensional
space corresponding to the twisted sector to which the bound state is
coupled to. The chiral multiplets  are charged under both gauge
fields, and represent the motion of a \emph{pair} of fractional branes
along one of the direction $z_i$. For instance, from eq.(\ref{sc1}) we
can see that $a_1,b_1$ are two chiral  fields charged with respect to
$A_\alpha^1$ and $A_\alpha^2$, so a pair of fractional branes of type
1 and 2 can move  along the first complex direction. The same holds
for a couple of fractional branes of type 3 and 4.

In string theory one can determine the properties of the fractional
Dp-branes of the orbifold $\mathbb{C}^3 / (\mathbb{Z}_2 \times
\mathbb{Z}_2)$ by computing the vacuum energy $Z$ of the open strings
stretched between two of them that is given by:
\begin{equation}
Z =  \int_{0}^{\infty} \frac{ds}{s} Tr_{NS-R} \left[\left(\frac{1 +
 (-1)^F}{2} \right) \left( \frac{e + h_1 +h_2 + h_3}{4}\right) {\rm
 e}^{-2 \pi s (L_0 -a)}  \right]
\label{zz87}
\end{equation}
where the first term under the trace performs the GSO projection, the
second  term the orbifold projection in the case of our orbifold,
while  $a =1/2$ in the NS sector and $a=0$ in the Ramond sector. The
properties  of the fractional D3-branes are easily studied by
performing in eq.(\ref{zz87}) the modular transformation $s
\rightarrow t =1/s$ that brings us to the closed string channel, and
by rewriting $Z$ as a matrix element between two boundary states with
the insertion of closed string propagator.  Since the closed string
theory living on our orbifold has an untwisted sector together with
three twisted sectors we have to consider a boundary state for each of
the previous four sectors. They are related to the four terms that
appear in the second bracket under the trace in eq.(\ref{zz87}). When
one takes the $e$ inside the bracket corresponding in the closed
string channel to the untwisted sector, one gets $1/4$ of the
contribution of the open strings  stretched between two D3-branes in
flat space. This means that the boundary state corresponding to the
untwisted sector is equal to the one in flat space apart from an
additional normalization factor $1/2$. The other three terms in the
second bracket in eq.(\ref{zz87}), when rewritten in the closed string
channel, will determine the boundary states corresponding to the three
twisted sectors.  Let us consider for instance the term denoted with
$h_1$ in the second bracket  in eq.(\ref{zz87}). Since $h_1$ (see
table 1) acts by changing  sign to $z_2$ and $z_3$, but leaving $z_1$
invariant, the twisted boundary state that we will obtain will be
equal to the one  corresponding to the twisted sector of the orbifold
$\mathbb{C}^2/\mathbb{Z}_2$, where $\mathbb{C}^2$ is spanned by $z_2$
and $z_3$, with an additional normalization factor $1/\sqrt{2}$ with
respect to the case of the orbifold $\mathbb{C}^2/\mathbb{Z}_2$. The
same is true for the other two terms $h_2$ and $h_3$ where
$\mathbb{C}^2$ is spanned respectively by $z_1$ and $z_3$ and by $z_1$
and $z_2$. In conclusion in the case of the orbifold $\mathbb{C}^3/
(\mathbb{Z}_2 \times \mathbb{Z}_2 )$ we can just take the boundary
states as given in \cite{antonella,Frau:2000gk} for the orbifold
$\mathbb{C}^2 /\mathbb{Z}_2$ and multiply them with the additional
normalization factor $1/\sqrt{2}$.

The previous considerations allow us to write almost immediately the
coupling of  the fractional D3-branes of the orbifold  $\mathbb{C}^3 /
(\mathbb{Z}_2 \times \mathbb{Z}_2)$ from those published in eqs. (7)
and (8) of \cite{Bertolini:2000dk} corresponding to the orbifold
$\mathbb{C}^2/\mathbb{Z}_2$. The supergravity fields a fractional
D3-brane couples to are the metric and the RR 4-form potential in the
untwisted sector and  3 scalars $b_i$ and 3 4-form potential $A_4^i$
in the twisted sector. The 3 scalars  and the 3 4-form potentials
correspond to the dimensional reduction of the Kalb-Ramond  2-form
potential $B_2$ and of the 6-form potential $C_6$ on the three  cycles
${\cal C}_i$, respectively (with charges according to
eqs.(\ref{ga1})-(\ref{ga4})). In doing that one should remember that
for our orbifold the background values of the $B_2$-fluxes
are~\cite{bflux}:
\begin{equation}
\label{bfb}
\int_{{\cal C}_i} B_2 = 4 \pi^2 \alpha' \frac{1}{2} \equiv b_0
\end{equation}
for any $i$, and $b_i \equiv \int_{{\cal C}_i} B_2 = b_0 + \tilde
b_i$, where $\tilde b_i$ are the fluctuations of the fluxes around
$b_0$.  The couplings of the untwisted fields with a fractional
D3-brane  are given by:
\begin{equation} 
\braket{B}{h} = -\frac{T_3}{4}\,\, h_{\alpha}^{\,\,\, \alpha}
\,V_4~~~,~~~ \braket{ B}{ C_{4}} = \frac{T_3}{4\,\kappa} \,C_{0123}\,
V_4
\label{untw9}
\end{equation}
where $T_3 =\sqrt{\pi}$ is the normalization of the boundary state
which is related to the brane tension in units of the gravitational
coupling constant $\kappa= 8 \pi^{7/2}(\alpha')^{2} g_{s}$ \cite{bs},
$V_4$ is the (infinite) world-volume of the D3-brane, and the index
$\alpha$ labels the longitudinal directions. The couplings of the
twisted fields with  a fractional D3-brane are given by:
\begin{equation} 
\braket{B}{\widetilde b_i} = -\frac{T_3}{4 \,\kappa}
\,\frac{1}{2\pi^2\alpha'}\,\widetilde b_i \, V_4 ~~~,~~~
\braket{B}{A_{4}^i} = \frac{T_3}{4 \,\kappa}\,\frac{1}{2\pi^2 \alpha
'}\,A_{0123}^i\, V_4
\label{twi86}
\end{equation}
From the previous couplings one can  deduce the boundary action of a
fractional D3-brane of any kind. For a  brane of type 1, which has
positive charge with respect to any twisted  sector, the boundary
action is given by:
\[
S_1 = - \frac{\tau_3}{4} \int d^{4} x \sqrt{- \det G_{\alpha \beta}}
\left[1 + \frac{1}{2 \pi^2 \alpha'} \sum_{i=1}^{3} {\tilde{b}}_i
\right] +
\]
\begin{equation}
+ \frac{\tau_3}{4} \int \left[C_4 \left(1 + \frac{1}{2 \pi^2 \alpha'}
\sum_{i=1}^{3} {\tilde{b}}_i \right) + \frac{1}{2 \pi^2 \alpha'}
\sum_{i=1}^{3} A_{4}^{i} \right]
\label{biac56}
\end{equation}
where $\tau_3 = \frac{T_3}{\kappa}$. For the other types of fractional
D3-branes the boundary action has the same structure, the only
difference are the signs in the coupling to the twisted sectors which
can be found in eqs.(\ref{ga1})-(\ref{ga4}).

From the above couplings one can also compute the large distance
behaviour of the various fields. For the metric one gets:
\begin{equation}
ds^2 \simeq \left(1-\frac{Q}{2\,r^4}\right)\,\eta_{\alpha\beta}
\,dx^\alpha dx^\beta + \left(1+\frac{Q}{2\,r^4}\right)\,\delta_{lm}
\,dx^ldx^m\,+...
\label{met1}
\end{equation} 
where $\alpha,\beta=0,...,3$; $l,m=4,...,9$;
$r=\sqrt{x^lx^m\delta_{lm}}~$ and $Q = \pi\, g_s\,(\alpha')^2$ while
for the untwisted 4-form potential one gets
\begin{equation} 
C_{4} \simeq -\,\frac{Q}{r^4}~dx^0\wedge dx^1\wedge dx^2\wedge dx^3
+...
\label{c41}
\end{equation}
The asymptotic behaviour of the twisted fields is instead given by
\begin{eqnarray}
{\widetilde b_i} &\simeq& K\,\log (\rho_{i}/\epsilon)+...
\label{b1} \\
A_{4}^{i} &\simeq& K\,\log (\rho_{i}/\epsilon)~dx^0\wedge dx^1\wedge
dx^2 \wedge dx^3 +...
\label{a41}
\end{eqnarray}
where $\rho_i = |z_i|$, $\epsilon$ is a regulator and
\begin{equation}
\label{kdef}
K = \frac{T_3}{4\kappa} \frac{1}{2\pi^2 \alpha'} \frac{2 \kappa^2
4}{2\pi}= 4\,\pi\,g_s\,\alpha'
\end{equation}
By considering fractional branes of type 2, 3 and 4 one gets similar
results, the only difference  being the sign of the coupling to the
twisted fields, according to eqs.(\ref{ga2})-(\ref{ga4}).
  
\vskip .7cm

%%%%%%%%%%%%%%%%%%%%%%%%%%%%%%%%%%%%%%%%%%%%%%%%%%%%%%%%%%%%%%%%%%%%%%%%%
%%%%%%%%%%%%%%%%%%%%%%%%%%%%%%%%%%%%%%%%%%%%%%%%%%%%%%%%%%%%%%%%%%%%%%%%%
\section{Classical solution for fractional D3-branes}
\label{section3}
In this section we derive the supergravity solution describing a bound
state  of $N_1$ fractional D3-branes of type 1, $N_2$ of type 2, $N_3$
of type 3 and  $N_4$ of type 4 in the orbifold theory  we are
considering. Let us start from the action of type IIB supergravity in
ten dimensions:
\begin{eqnarray}
&&S_{IIB}= \frac{1}{2 \kappa^2 }\left\{ \int  d^{10} x
\sqrt{-\mbox{det} G} \, R - \frac{1}{2} \int \left[ d\phi \wedge
*d\phi\, +\, \mbox{e}^{-\phi} H_3\wedge *H_3 \, + \mbox{e}^{2
\phi}F_1\wedge *F_1\right.\right.\nonumber\\ &&\left. \left.+
\mbox{e}^{\phi} \tilde{F}_3\wedge *\tilde{F}_3 \, + \frac{1}{2}
\tilde{F}_5\wedge *\tilde{F}_5 - C_4 \wedge H_3 \wedge
F_3\right]\right\}
\label{action}
\end{eqnarray}
where
\begin{equation}
H_3=dB_2\,\,\,,\,\,\, F_1=dC_2\,\,\,,\,\,\,F_3=dC_2\,\,\,,\,\,\, F_5
=dC_4
\label{deffst}
\end{equation}
are, respectively the field strengths of the Kalb-Ramond two form and
the 0-,2- and 4-form RR potentials, and
\begin{equation}
\tilde{F}_3\, =\, F_3 \,+\, C_0 \wedge H_3 \,\,\,\,\,,\,\,\,\,\,
\tilde{F}_5=F_5\, +\, C_2\wedge H_3
\label{deffs1}
\end{equation}
As usual the self-duality constraint $*\tilde{F}_5=\tilde{F}_5$ has to
be implemented on shell.

Since we are interested in computing the classical solution of
fractional D3-branes in the orbifold $\mathbb{C}^3/(\mathbb{Z}_2
\times \mathbb{Z}_2)$, it is convenient to introduce the complex
fields:
\begin{equation}
\tau=C_0 + i e^{-\phi} \quad \mbox{and}\quad G_{3}= dC_2+\tau dB_2
\label{fieldcom}
\end{equation}
and the standard D3-brane ansatz for the untwisted fields $G_{MN}$ and
$\tilde{F}_5$, namely:
\begin{eqnarray}
&&ds^2= H^{-1/2}\eta_{\alpha\beta}\,dx^\alpha \,dx^\beta +
H^{1/2}\delta_{lm}\,dx^l\, dx^m
\label{metric01}\\
&&\tilde F_5= dH^{-1}\wedge V_4+*\left( d H^{-1}\wedge V_4\right)
\label{f_5}
\end{eqnarray}
In terms of the complex fields in eq.(\ref{fieldcom}), the equations
of motion for the axion and dilaton become:
\begin{equation}
d *\,d \tau + i {\rm e}^{\phi} d \tau \wedge *\, d \tau + \frac{i}{2}
G_3 \wedge *\, G_3 = 0
\label{dileq01}
\end{equation}
where, since the source we are interested in, namely fractional
D3-branes, does not couple to the dilaton and the axion, there is not
any source term in the right hand side of the above  equation. Hence,
requiring the above equation to be solved by constant dilaton and
axion, one gets back the constraint $G_3 \wedge *\,G_3=0$. Noticing
that
\begin{equation}
*\,G_3 = - \,H^{-1} \hat{*}_6 \,G_3 \wedge V_4
\end{equation}
where $\hat{*}_6$ depends only on the 6 transverse directions to the
D3-brane  and we have extracted all the warp factors, one can solve
the constraint imposed by the scalar equation by requiring:
\begin{equation}
\hat{*}_6 \, G_3 = - i \, G_3
\label{dileq03}
\end{equation}
The constant $i$ on the left hand side, instead, has been fixed by
observing that $\hat{*}_6\hat{*}_6 \, G_3 = - G_3$.
Eq.(\ref{dileq03}) seems a general condition satisfied by any
classical solution generated by  fractional branes living both on
orbifold and conifold geometry and it is related  to the supersymmetry
properties of the system \cite{GRANA,GUB,cve}.

The equations of motion for the two 2-forms can be grouped together in
the following equation:
\begin{equation}
d * G_3 + d \tau \wedge \left[i {\rm e}^{\phi} * \, G_3 + *\, H_3
\right] - i {\tilde{F}}_5 \wedge G_3 = - 2i \kappa^2 \left[
\frac{\delta  {\cal L}_b} {\delta B_2} - \tau \frac{\delta {\cal L}_b}
{\delta C_2}\right]
\label{g3eq01}
\end{equation}
where ${\cal L}_b$ is the Lagrangian density of the given source. We
now solve eq.(\ref{g3eq01}) by using the ansatz for the untwisted
fields given in eqs.(\ref{metric01}) and (\ref{f_5}), with constant
dilaton and axion, as already noticed. By plugging then
eq.(\ref{dileq03}) in eq.(\ref{g3eq01}) we get:
\begin{equation}
H^{-1} d \, \hat{*}_6 \, G_3 \wedge V_4 =  2i \kappa^2 \left[
\frac{\delta  {\cal {L}}_b}{\delta B_2} - \tau \frac{\delta {\cal
{L}}_b}{\delta C_2} \right]
\label{g3eq02}
\end{equation}
In order to solve the previous equation we have to write an ansatz for
the twisted fields, too. In this orbifold, as discussed in the
previous section, there are four kinds of fractional branes, each of
them coupled with all the twisted fields as it emerges from the linear
couplings dictated by the boundary state. We want to find a classical
background generated by a bound state made of all four different kinds
of fractional branes. A natural ansatz compatible with the large
distance behaviour of the twisted fields given in eqs.(\ref{b1}) and
(\ref{a41}) is:
\begin{equation}
G_3= d\gamma_{i} \wedge \omega_2^i
\label{g3eq03}
\end{equation}
with $i=1,2,3$, $\gamma_i=c_i+i b_i$ and where $c_i= \int_{{\cal C}_i}
C_2$ are the Hodge duals of the 4-form potentials the fractional
D3-branes actually couple to. Since the twisted fields $\gamma_i$ are
obtained by reducing $G_3$ along the cycles ${\cal C}_i$, they can
only  depend on the coordinates $z_i$. In particular, inserting the
above ansatz in eq.(\ref{dileq03}) one gets that $\gamma_i$ are
analytic functions of $z_i$. Moreover, by plugging the ansatz
(\ref{g3eq03}) in eq.(\ref{g3eq02}) we get, for each twisted
component, the following equation:
\begin{equation}
\delta^{rs} \partial_r \partial_s \gamma_i - 2 \pi i \, K f_i(N_I)\,
\delta(x^{2i+2})\, \delta(x^{2i+3} ) = 0
\label{g3eq04}
\end{equation}
where $ r,s \in \left\{ 2i+2, 2i+3\right\}$, $K$ is defined in
eq.(\ref{kdef})  while the functions $f_i(N_I)$ depend on numbers
$N_I$ of fractional branes of  the four different types and are given
by:
\begin{eqnarray}
&& f_1(N_I)= N_1 + N_2 - N_3 - N_4\nonumber\\  && f_2(N_I)= N_1 - N_2
+ N_3 - N_4 \nonumber\\  && f_3(N_I)= N_1 - N_2 - N_3 + N_4
\label{coupling}
\end{eqnarray}
The different signs in the previous expressions are due to the signs
appearing in the irreducible representations given in
eqs.(\ref{ga1})-(\ref{ga4}),  each of them corresponding to a
fractional brane of a given type.

One can easily see that the analytic solution of eq.(\ref{g3eq04}) is:
\begin{equation}
\gamma_i = i K \left[\frac{\pi}{2 g_s}+  f_i(N_I)\,\log
(z_i/\epsilon)\right]
\label{g3eq05}
\end{equation} 
where the background value given in eq.(\ref{bfb}) has been
introduced.  Let us now consider the field equation for the untwisted
4-form $C_4$ which  in this case looks like:
\begin{equation}
d*\tilde{F}_5\, - \,\frac{i}{2}\, G_3 \wedge \bar{G}_3\, + 2 \,
\kappa^2 \frac{\delta {\cal L}_b}{\delta C_4}=0
\label{eqf5}
\end{equation}
and which determines the warp factor $H$. Inserting in this equation
the Ansatz (\ref{metric01}), (\ref{f_5}) and (\ref{g3eq03}), we get:
\begin{equation}
\delta^{lm} \partial_l \partial_m H \,+\, \frac{1}{4}\sum_i
\left|\partial_{z_i}\gamma_i\right|^2 \delta^4_i(x) \,+\, 4 \pi^3 Q
\, f_0(N_I) \delta(x^4)\dots\delta(x^9)=0
\label{eqf51}
\end{equation}
where $f_0(N_I)= N_1 + N_2 + N_3 + N_4$ and $Q$ is defined before
eq.(\ref{c41}). The last equation is a generalization of the
corresponding one for fractional D3-branes in the orbifold
$\mathbb{C}^2/\mathbb{Z}_2$.  In this case we have three, instead of
one, terms depending on the twisted fields, and by plugging the result
(\ref{g3eq05})  in eq.(\ref{eqf51}), the solution will be just a
triple copy of the solution found in \cite{Bertolini:2000dk}:
\begin{equation}
H(r,z_i)= 1\,+\, f_0(N_I)\, \frac{Q}{r^4}+ \frac{K^2}{4\, r^4} \sum_i
f_i(N_I)^2 \left[ \log \left( \frac{r^4}{\epsilon^2(r^2-\rho^2_i)
}\right) -1 + \frac{\rho^2_i}{r^2-\rho^2_i} \right]
\label{sol01}
\end{equation}
with $\rho_i= |z_i|$. One can finally check that our solution also
satisfies the equation of motion for the metric. As a consistency
check  one can verify that the above solution does reproduce the
expected  asymptotic behaviour (\ref{met1})-(\ref{a41}).

From eq.(\ref{sol01}) one can see that the metric has a singularity of
repulson type \cite{kal}. This is quite a general feature of
supergravity solutions generated by non-conformal sources. The
singularity shows up because of the presence of the $K$-dependent term
in the function $H$, which is related to the coupling to the twisted
fields. This coupling is absent in the case of regular branes, which,
as discussed in section \ref{gspec}, are conformal. In ${\cal N}=2$
theories these singularities are often cured by an enhan\c{c}on
mechanism \cite{enhanc}, while for the ${\cal N}=1$ conifold case
discussed in \cite{KLEBA3} is the deformation of the conifold which
gives back a singularity free solution. As we will show in the next
section, in the present case it seems that an enhan\c{c}on phenomenon
is at hand, too, in agreement with observations recently made in
\cite{Merlatti:2001gd}. The enhan\c{c}on is a scale where new light
string degrees of freedom become relevant, due to fractional D-strings
becoming tensionless, in type IIB. This makes the supergravity action
one has started with, unable to describe the physics at scales smaller
than the enhan\c{c}on. In principle, by including these extra degrees
of freedom in the low energy action one should get back an
enhan\c{c}on free and singularity free solution, as discussed recently
in \cite{Wijnholt:2001us}. This is what we expect to be the case in
the ${\cal N}=1$ situation we are discussing, too, although we will
not address this problem here. In the next section, by doing a probe
analysis we show that fractional brane probes become tensionless at
the enhan\c{c}on. This excises the unwanted singularity, and is enough
for our present purpose, but at the same time limits the validity of
our solution to distances bigger than the enhan\c{c}on radius.
It would be interesting to  understand the relation between this
${\cal N}=1$ version of the  enhan\c{c}on phenomenon, with the pure
supergravity analysis performed  in \cite{KLEBA3}.

%%%%%%%%%%%%%%%%%%%%%%%%%%%%%%%%%%%%%%%%%%%%%%%%%%%%%%%%%%%%%%%%%%%%%%%%%
%%%%%%%%%%%%%%%%%%%%%%%%%%%%%%%%%%%%%%%%%%%%%%%%%%%%%%%%%%%%%%%%%%%%%%%%%
\section{Non-conformal ${\cal N}=1$ SYM and brane probe}
\label{probe}

According to the discussion in section \ref{gspec}, we will now
consider  the low energy dynamics of a bound state of, say, $N_1$
branes of type 1  and $N_2$ branes of type 2 (all what follows can be
applied to any couple  of fractional brane types). The supergravity
background we are going to probe is then  the one discussed in section
\ref{section3} with $N_3=N_4=0$. Making a probe analysis  we will see
that supergravity can predict the Wilsonian $\beta$-function of the
corresponding  gauge theory \cite{Herzog:2001xk} (for a  review on the
probe technique we recommend \cite{joh}).

Before doing that, let us fix once for all the relation between the
gauge couplings of the four possible gauge groups and the fluxes of
the NS-NS 2-form $B_2$ along the  three shrinking spheres . This can
be easily done by considering, for any given type of fractional brane,
the gauge kinetic term arising from the DBI action  when a $U(1)$
$F_{\alpha\beta}$ field is switched-on on the world-volume. Let us
consider, for instance, a brane of type 1. Its DBI action, which we
have derived in section \ref{gspec}, can be equivalently written as
\footnote{For the sake of simplicity in this formula (and subsequent
ones) we have introduced the dimensionless field $\hat B_2$ which is
related to $B_2$ as $\hat B_2 = (4\pi^2\alpha')^{-1} B_2$.}
\begin{eqnarray}
S &=& - \frac{T_3}{2\kappa} \int d^4x\,\sqrt{-\det \left( G + 2 \pi
\alpha' F \right)_{\alpha\beta}} \left(\int_{{\cal C}_1} \hat B_{2} +
\int_{{\cal C}_2} \hat B_{2} + \int_{{\cal C}_3} \hat B_{2} - 1 \right)
\end{eqnarray}
Expanding the above action  up to the quadratic terms in the gauge
field one gets (recall the generators  are normalized as
$Tr(T_aT_b)=1/2 \,\delta_{ab}$):
\begin{eqnarray}
S &=& - \frac{T_3}{4\,\kappa}\,(2 \pi \alpha')^2\,\int d^4x\,
\frac{1}{4}F_{\alpha\beta} F^{\alpha\beta} \left(\int_{{\cal C}_1}
\hat B_{2} + \int_{{\cal C}_2} \hat B_{2} + \int_{{\cal C}_3} \hat
B_{2} - 1 \right) + ...
\end{eqnarray}
By substituting $\frac{T_3}{4\,\kappa}\,(2 \pi \alpha')^2  = 1/(8\pi
g_s)$ one finally gets for the gauge coupling:
\begin{eqnarray}
\label{gc11}
\frac{1}{g_1^2} =  \frac{1}{8\pi g_s}\,\left(\int_{{\cal C}_1} \hat
B_{2} + \int_{{\cal C}_2} \hat B_{2} + \int_{{\cal C}_3} \hat B_{2}  -
1 \right)
\end{eqnarray}
Repeating the above reasoning for all other kinds of fractional branes
one ends up with the following set of relations:
\begin{eqnarray}
\label{gc1}
\frac{1}{g_1^2}&=&\frac{1}{8\pi g_s}\left( \int_{{\cal C}_1} \hat
B_{2} + \int_{{\cal C}_2} \hat B_{2} + \int_{{\cal C}_3} \hat B_{2} -
1 \right) \\
\label{gc2}
\frac{1}{g_{2}^2}&=& \frac{1}{8\pi g_s}\left(1 + \int_{{\cal C}_1}
\hat B_{2} - \int_{{\cal C}_2} \hat B_{2} -  \int_{{\cal C}_3} \hat
B_{2}\right) \\
\label{gc3}
\frac{1}{g_{3}^2}&=&\frac{1}{8\pi g_s}\left(1 - \int_{{\cal C}_1} \hat
B_{2} + \int_{{\cal C}_2} \hat B_{2} - \int_{{\cal C}_3} \hat
B_{2}\right) \\
\label{gc4}
\frac{1}{g_{4}^2}&=&\frac{1}{8\pi g_s}\left(1 - \int_{{\cal C}_1} \hat
B_{2}- \int_{{\cal C}_2} \hat B_{2} + \int_{{\cal C}_3} \hat
B_{2}\right)
\end{eqnarray}
These are the master formul\ae $\,$ relating gauge theory parameters
(left hand side)  with supergravity fluxes (right hand side). One
should remember, however (as already discussed in the Introduction),
that the probe analysis, in relating the energy scale of the gauge
theory to some transverse length in the supergravity solution (for
spherically symmetric solutions  $\Lambda = (2\pi\alpha')^{-1} r$
\cite{Peet:1998wn}), implies that there should be some scalar field
acquiring a v.e.v. in the effective gauge theory one is
describing. This  is because transverse directions are seen as scalar
fields on the D-brane. Therefore,  while the above formul\ae $\,$ are
indeed correct, they cannot provide a probe analysis prediction for
the pure ${\cal N}=1$ super Yang-Mills theory since, in that case,
there are no scalars  to relate the energy with. For this reason one
should use composite probes to test the supergravity background. We
will reconsider the case of the pure ${\cal N}=1$ theory at the end of
the section.

Given the above general formul\ae, let us now come back to the
analysis of the bound state we want to probe.  The effective gauge
theory living on $N_1$ branes  of type 1 and $N_2$ branes of type 2 is
a ${\cal N}=1$ super Yang-Mills with gauge group $U(N_1)\times
U(N_2)$. The diagonal $U(1)$ factor is free and the relative $U(1)$
factor is subleading to first order in $1/N_1$ and $1/N_2$. Here we
will be mainly concerned with the running of the couplings $g_1$ and
$g_2$ for the semi-simple factor $SU(N_1)\times SU(N_2)$.  In addition
to the gauge multiplets, we have two chiral multiplets, one
transforming in the $(N_1,\bar N_2)$ and the other in  the $(\bar
N_1,N_2)$. The two (Wilsonian) $\beta$-functions are
\begin{eqnarray}
\label{beta12}
\beta(g_1) &=& \frac{g_1^3}{(4\pi)^2}  \left[\underbrace{-\frac{11}{3}
N_1 + \frac{2}{3} N_1}_{Gauge\;multiplet} + \underbrace{\frac{1}{6} 2
N_2 +  \frac{1}{3} 2 N_2}_{Chiral\;multiplet} \right] =   -
\frac{g_1^3}{(4\pi)^2}\, (3 N_1 - N_2) \\ \beta(g_2) &=&
\frac{g_2^3}{(4\pi)^2}  \left[\underbrace{-\frac{11}{3} N_2 +
\frac{2}{3} N_2}_{Gauge\;multiplet} + \underbrace{\frac{1}{6} 2 N_1 +
\frac{1}{3} 2 N_1}_{Chiral\;multiplet} \right] =   -
\frac{g_2^3}{(4\pi)^2}\, (3 N_2 - N_1) \\
\end{eqnarray}
and the corresponding gauge couplings are
\begin{eqnarray}
\label{g1}
\frac{1}{g_1^2} &=& \frac{1}{g_{0}^2}  \left( 1+
\frac{g_{0}^2}{4\pi^2} \frac{3 N_1 -N_2}{2} \log\mu\right) \\
\label{g2}
\frac{1}{g_2^2} &=& \frac{1}{g_{0}^2}  \left( 1+
\frac{g_{0}^2}{4\pi^2} \frac{3 N_2 -N_1}{2} \log\mu\right)
\end{eqnarray}
where $g_0$ is the bare gauge coupling which can be assumed to be the
same for both groups without loss of generality, since it drops out of
the $\beta$-function. To obtain the above gauge quantities from
supergravity by probe analysis, one has to start from a bound state of
$N_1+1$ and $N_2+1$ fractional branes of type 1 and 2, respectively,
corresponding to  $U(N_1+1)\times U(N_2+1)$ gauge theory.

We have seen that it is impossible to move a single brane from the
orbifold point while it is possible to move a pair of branes of type 1
and 2 (``the probe''). This corresponds to ``Higgsing'' the gauge
theory in  a certain way. We shall now discuss this phenomenon in
detail  from the gauge theory point of view and show how one can
relate the (frozen) coupling that one reads on the probe using the
supergravity analysis to the (running) couplings of the
$SU(N_1+1)\times SU(N_2+1)$ theory at the gauge symmetry breaking
point.

In order to avoid unnecessary complications, we shall consider the
$U(N_1+1)\times U(N_2+1)$ theory and neglect the contribution from the
relative $U(1)$ field, which is subleading to first order in $1/N_1$
and $1/N_2$. The breaking
\begin{equation}
  U(N_1+1)\times U(N_2+1) \rightarrow U(1)^\prime \times U(N_1) \times
  U(N_2)
\end{equation}
where the group $U(1)^\prime$ refers to the gauge field on the probe,
is accomplished  by giving the following v.e.v. to the scalar
components of the chiral multiplet  $\Phi_1$:
\begin{equation}
a_1 = b_1^\mathrm{T} = \begin{pmatrix} v  & 0  & \cdots & 0  \cr 0 & 0
                                           & \cdots & 0 \cr \vdots &
                                           \vdots & & \vdots \cr 0 & 0
                                           & \cdots & 0 \cr
                                           \end{pmatrix},  \label{flat}
\end{equation}
where $a_1$ is a $N_1 \times N_2$ matrix. We can assume that $v$ is
real  without loss of generality. Equation (\ref{flat}) represents a
classical flat direction and corresponds to moving a pair of branes of
type 1 and 2 away from the orbifold point in the $z_1$ direction.

If we write the gauge fields $A_1$ and $A_2$ corresponding to
$U(N_1+1)$ and $U(N_2+1)$ respectively as square matrices\footnote{In
the following we shall always suppress the Lorentz index $\alpha$ on
the gauge field.}  it is clear that the gauge bosons that become
massive are those in the first row and first column. More explicitly,
if we write
\begin{equation}
    A_1 =\frac{1}{\sqrt{2}}  \begin{pmatrix}  A_1^0 & W_1^1  & \cdots
            & W_1^{N_1}  \cr W_1^{1*}  &        &        & \cr \vdots
            &        &        &            \cr W_1^{N_1 *}  &        &
            &            \cr \end{pmatrix},\;\; A_2 = -
            \frac{1}{\sqrt{2}}  \begin{pmatrix}  A_2^0 & W_2^1 &
            \cdots & W_2^{N_2}  \cr W_2^{1*}  &        &        & \cr
            \vdots   &        &        &            \cr W_2^{N_2 *}  &
            &        &            \cr \end{pmatrix}
\end{equation}
we see that all $W$ fields become massive and so does the linear
combination $\propto g_1 A_1^0 + g_2 A_2^0$, whereas the linear
combination  $\propto g_2 A_1^0 - g_1 A_2^0$ remains massless. In the
same way, all the chiral fields in the first  rows and columns of
$a_1$ and $b_1$ are ``eaten'' by the massive gauge multiplets except
for one linear combination representing the motion of the probe. This
chiral multiplet is not charged with respect to  $g_2 A_1^0 - g_1
A_2^0$ and thus the theory on the probe becomes  free and its coupling
constant $g$ stops running.  Thus, by reading off the value of such
coupling in the IR we obtain information about the value of the
couplings $g_1$ and $g_2$ at the breaking point.

To obtain the exact formula, we need to normalize the fields on the
brane. The correct normalization is:
\begin{equation}
     Z = \frac{1}{\sqrt{g_1^2 + g_2^2}} \left(g_1 A_1^0 + g_2
         A_2^0\right)\quad\hbox{and}\quad \gamma =
         \frac{1}{\sqrt{g_1^2 + g_2^2}} \left(g_2 A_1^0 - g_1
         A_2^0\right)
\end{equation}
where $Z$ and $\gamma$ are the massive and massless bosons
respectively. The  relation between the probe coupling constant $g$
with $g_1$ and $g_2$ becomes thus:
\begin{equation}
\label{gexp}
\frac{1}{g^2} = \frac{1}{g_1^2} + \frac{1}{g_2^2} = \frac{2}{g_0^2} +
\frac{1}{4\pi^2} (N_1 + N_2) \log\mu
\end{equation}
where, in doing the last step, we have used eqs.(\ref{g1}) and
(\ref{g2}).

Let us now consider the (probe) action for a bound state of a
fractional  brane of type $1$ and a fractional brane of type $2$. The
former is\footnote{We use for the R-R twisted fields $A^i_{(4)}$ the
same convention we  have introduced for the 2-form $B_2$.}
\begin{eqnarray}
S_{1} &=& - \frac{T_3}{2\kappa}\left[ \int d^4x\,\sqrt{-\det
G_{\alpha\beta}} \left(\int_{{\cal C}_1} \hat B_{2} +  \int_{{\cal
C}_2} \hat B_{2} + \int_{{\cal C}_3} \hat B_{2} - 1 \right)\right] +
\nonumber \\ &+&\frac{T_3}{2\kappa} \left[C_{(4)}\left(\int_{{\cal
C}_1} \hat B_{2} + \int_{{\cal C}_2} \hat B_{2} + \int_{{\cal C}_3}
\hat B_{2} - 1 \right) -  \, \int \sum_{i=1}^3 \hat A^i_{4}\right]
\end{eqnarray}
while the latter is
\begin{eqnarray}
S_{2} &=& - \frac{T_3}{2\kappa}\left[ \int d^4x\,\sqrt{-\det
G_{\alpha\beta}} \left( 1 + \int_{{\cal C}_1} \hat B_{2} - \int_{{\cal
C}_2} \hat B_{2} - \int_{{\cal C}_3} \hat B_{2} \right) \right] +
\nonumber \\  &+& \frac{T_3}{2\kappa}\left[\int C_{(4)}\left( 1 +
\int_{{\cal C}_1} \hat B_{2} - \int_{{\cal C}_2} \hat B_{2} -
\int_{{\cal C}_3} \hat B_{2}\right) + \,   \int \left(\hat A^1_{4} -
\hat A^2_{4} - \hat A^3_{4}\right)\right] \nonumber \\
\end{eqnarray}
By summing them up one obtains
\begin{eqnarray}
\label{bsum}
S_{1+2} &=& - \frac{T_3}{2\kappa}\left[ \int d^4x\,\sqrt{-\det
G_{\alpha\beta}} \int_{{\cal C}_1} 2 \,\hat B_{2} -  \int C_{4}
\int_{{\cal C}_1} 2\, \hat B_{2} - \, \int 2 \,\hat A^1_{4} \right]
\end{eqnarray}
where the coupling to the last two twisted sectors has cancelled and
the twisted fields left depend on $z_1$, only. The system is then free
to move in the  $x_4,x_5$ plane and probe computations are allowed, as
anticipated.

In order to find the effective gauge coupling describing the low
energy dynamics  of the probe, one can simply repeat a reasoning
similar to that at the beginning of the section.  Indeed, repeating
the DBI action expansion  for the gauge field kinetic term  previously
described, one gets for the probe gauge coupling $g$:
\begin{equation}
\label{gcg}
\frac{1}{g^2} =  \frac{2}{8 \pi g_s} \int_{{\cal C}_1} \hat B_{2} =
\frac{1}{8 \pi g_s}  + \frac{1}{4 \pi^2} (N_1 + N_2) \log \mu_1
\end{equation}
where we have inserted the classical solution given in eq.(\ref{b1}).
From the above expression we get the corresponding (Wilsonian)
$\beta$-function to be:
\begin{equation}
\label{bg}
\beta = \mu_1 \frac{\partial g}{\partial \mu_1} = -  \frac{ (8\pi
g_s)^{1/2}}{\left[1 +\frac{8 \pi g_s}{4 \pi^2}\frac{N_1 + N_2}{2} \log
\mu \right]^{3/2}}\; \frac{g_0^2 (N_1 +N_2)}{16 \pi^2} = - \frac{2(N_1
+ N_2)}{(4\pi)^2}\,g^3
\end{equation}
This is the correct result expected from eqs. (\ref{gc1}) and
(\ref{gc2}) which indeed imply that:
\begin{eqnarray}
\label{sumg}
\frac{1}{g_1^2} + \frac{1}{g_2^2} &=& \frac{2}{8 \pi g_s} \int_{{\cal
C}_1} \hat B_{(2)} \,= \, \frac{1}{g^2}
\end{eqnarray}
This coincides with the result from gauge theory found in
eq.(\ref{gexp})!   As one can see, the supergravity prediction,
eq.(\ref{gcg}), is in precise  (numerical) agreement with the above
equation and eqs.(\ref{g1}) and (\ref{g2}). Notice that in
eq.(\ref{gexp}) one has  $g_0^2= 16 \pi g_s$, this being consistent
with eqs.(\ref{g1})-(\ref{g2})  and (\ref{gc1})-(\ref{gc2}).

As anticipated in the previous section, an enhan\c{c}on phenomenon
seems at hand here. Indeed  the probe becomes tensionless at a
distance $\hat \rho_1$, the enhan\c{c}on, given by:
\begin{equation}
\hat \rho_1 = \epsilon \; e^{- \pi / 2 (N_1 + N_2) g_s}
\end{equation}
This excises the unwanted repulson singularity from the solution,
since it indicates the appearance of new light degrees of freedom
which are expected to become relevant at the enhan\c{c}on scale and to
affect the low energy physics. At the same time, the vanishing of  the
probe at $\hat \rho_1$ makes the geometry at distances $\rho < \hat
\rho_1$ out of reach. All gauge theory information are then confined
to the perturbative region, as it is usually the case for situations
in which an enhan\c{c}on locus shows-up in the geometry. Supergravity
alone, at least using probe techniques, seems not able to give
information on the non-perturbative region of the gauge theory. To  go
further one should include more states, as recently discussed in
\cite{Wijnholt:2001us}.

Eq.(\ref{sumg}) gives the correct gauge theory prediction for any
value of $N_1,N_2$.  Since we are working with a perturbatively
renormalizable theory in the UV,  where the anomalous dimensions
$\gamma$ are small, it is possible for the two  $\beta$-functions to
be both UV-free (it is sufficient that $1/3 N_1 < N_2 < 3 N_1$). Also,
one can make one of the gauge group conformal and the other running by
choosing $N_1=3N_2$. In this way $\beta(g_2)=0$ and one directly gets
the $\beta$-function for $g_1$ from that of $g$.  This has to be
compared to the case discussed in \cite{KLEBA2,KLEBA3,Herzog:2001xk},
which deals with the IR behaviour of the  theory away from the
gaussian fixed point, where  $\gamma \approx -1/2$. In that case, for
any choice of $N_1, \; N_2$, the two couplings will run in  opposite
directions.
 
As we have already discussed, we cannot really probe the pure ${\cal
N}=1$  super Yang-Mills with the probe technique. Nevertheless, let us
notice the following  fact. Consider, for instance, a supergravity
background with just one kind of fractional D3-branes, i.e. $N_1$
branes of type 1, which are described at low energy by pure ${\cal
N}=1$  super Yang-Mills with gauge group $U(N_1)$. Plugging the value
for the $B_2$-fluxes dictated by the corresponding supergravity
solution in formula  (\ref{gc11}) with $\rho_1=\rho_2=\rho_3\equiv
\rho$, one gets precisely the gauge coupling and the Wilsonian
$\beta$-function of  ${\cal N}=1$ pure super Yang-Mills, namely:
\begin{eqnarray}
\frac{1}{g_1^2}&=&\frac{1}{8\pi g_s}\left( \int_{{\cal C}_1} \hat
B_{2} + \int_{{\cal C}_2} \hat B_{2} + \int_{{\cal C}_3} \hat B_{2} -
1\right) =  \frac{1}{g_{0}^2}  \left( 1+ \frac{g_{0}^2}{4\pi^2}
\frac{3 N_1}{2}  \log\mu\right)
\end{eqnarray}
and
\begin{equation}
\beta(g_1) = - \frac{3 N_1}{(4\pi)^2}\,g_1^3
\end{equation}
Physically this probe analysis cannot really be done since the probe
does  not have any moduli associated to it, and is stuck at the
orbifold fixed point. This is of course a region  which is out of
reach of the supergravity solution, because corresponds   to distances
smaller then the repulson singularity. It is however worth noticing
how the matching holds in this case, too. Perhaps this result can be
justified by a refinement of the  probe analysis.

%%%%%%%%%%%%%%%%%%%%%%%%%%%%%%%%%%%%%%%%%%%%%%%%%%%%%%%%%%%%%%%%%%%%%%%%%%%%
\vskip 0.8cm
\noindent
{\large {\bf Acknowledgments}}
\vskip 0.2cm
\noindent
We thank M. Bill\`o for partecipating in the early stage of this work
and for many discussions, and  M. Frau, E. Imeroni, A. Lerda,
E. Lozano-Tellechea, B.E.W. Nilsson, F. Roose, R. Russo,
P. Salomonson, F. Sannino and D. Tsimpis for useful
discussions. G.F. and R.M. would like to thank NORDITA for the kind
hospitality. This work is partially supported by the EC  RTN contracts
HPRN-CT-2000-00131 and HPRN-CT-2000-00122. M.B. is supported by  a EC
Marie Curie Postdoc Fellowship under contract number
HPMF-CT-2000-00847.

%%%%%%%%%%%%%%%%%%%%%%%%%%%%%%%%%%%%%%%%%%%%%%%%%%%%%%%%%%%%%%%%%%%%%%%%%%%%
%%%%%%%%%%%%%%%%%%%%%%%%%%%%%%%%%%%%%%%%%%%%%%%%%%%%%%%%%%%%%%%%%%%%%%%%%%%%

\end{document}